\documentclass[aps,pre,twocolumn,amsfonts,amssymb]{revtex4-1}

\usepackage{amsmath}
\usepackage{amssymb}
\usepackage{amsfonts}
\usepackage{graphicx}

\begin{document}

\title{Manifestation of T-Exciton Migration in the Kinetics of
Singlet Fission in Organic Semiconductors}

\author{A. I. Shushin }

\affiliation{Institute of Chemical Physics, Russian Academy of
Sciences, 119991, GSP-1, Kosygin Str. 4, Moscow, Russia}

\begin{abstract}
Kinetics of singlet fission in organic semiconductors, in which the
excited singlet state (${\rm S}_1$) spontaneously splits into a pair
of triplet (T) excitons, is known to be strongly influenced by back
geminate annihilation of TT-pairs. We show that this influence can
be properly described only by taking into account the diffusive
exciton migration. The migration effect is treated in the model of
two kinetically coupled states: the intermediate state of
interacting TT-pairs and the state of migrating excitons. Within
this model the singlet fission (including magnetic field effects) is
studied as applied to the fluorescence decay kinetics (FDK) ${
I}_{_{{\rm S}_1}}\!(t)$ for ${\rm S}_1$-state. The analysis shows
that migration strongly affects the FDK resulting, in particular, in
the universal long-time dependence ${ I}_{_{{\rm S}_1}}\!(t) \sim
t^{-3/2}$. The model accurately describes the FDK, recently observed
for a number of systems. Possible applications of the considered
model to the analysis of mechanisms of migration, using
experimentally measured FDK, are briefly discussed.
\end{abstract}

\maketitle

\bigskip

\section{Introduction}

Singlet fission, i.e. spontaneous splitting of the optically excited
singlet state ${{\rm S}_{_{\!}1\!}^{^{\!{*}}}}$ into a pair of
triplet (T) excitons (TT-pair), is the important photophysical
process, playing the key role in many phenomena, which significantly
control photovoltaic and spintronic properties of organic
semiconductors, important for applications \cite{Mi1,Sw1,Zu}. This
process
is actively investigated for tens of years \cite{Mi1,Sw1,Sw2}.
Intensive experimental investigations of fission kinetics inspired
considerable theoretical studies of this process
\cite{Mi1,Sw1,Sw2,Mer1,Su,Shu00,Shu01,Ber1,Bi1,Bi2}.

Specific features of singlet fission are usually analyzed within the
model represented by the kinetic scheme
\begin{equation} \label{form0}
{{\rm S}_{0}^{}+{\rm S}_{0}^{}}
\stackrel{\,\,^{k_r^{}}}{\leftarrow}
({{\rm S}_{0}^{}+{\rm S}_{_{\!}1\!}^{^{\!{*}}}})\,
\rightleftarrows_{\!\!\!\!\!\!_{_{\hat K_{\!s}}}}
^{\!\!\!\!\!\!\!\!^{^{k_{-\!s}}}}\, [{\rm TT}]\,
\rightleftarrows_{\!\!\!\!\!\!\!_{_{k_{-\!e}}}}
^{\!\!\!\!\!\!^{^{k_{e}}}}\, [{\rm T\!+\!T}],
\end{equation}
in which all stages are conventionally treated as first order
reactions. The primary stage of the fission process is the
transition (with the rate $k_{-s}^{}$) from the initially excited
state $({\rm S}_{0}^{}+{{\rm S}_{_{\!}1\!}^{^{\!{*}}}})$ into the
intermediate [TT]-state of coupled T-excitons (called also as
$c$-state). Evolution of [TT]-state is determined by geminate
T-exciton annihilation, dissociation into a pair of separate
T-excitons, denoted as [T+T]-state (or $e$-state), and back
(geminate) capture into [TT]-state with rates $\hat K_{s}^{}$,
$\,k_{e}^{}$ and $k_{-e}^{}$, respectively. Note that
TT-annihilation is a spin-selective process (with the rate $\hat
K_{s}^{}$ depends on the total TT-spin ${\bf S} = {\bf S}_{a}^{} +
{\bf S}_{b}^{}$ [eq. (\ref{form4})]), which leads to the dependence
of singlet fission kinetics on the magnetic field $B$ (see below).

The fission process is accompanied by deactivation of ${{\rm
S}_{_{\!}1\!}^{^{\!{*}}}}$-state with the total rate $k_r^{}$,
resulting from radiative and non-radiative transitions with rates
$\kappa_r^{}$ and $\kappa_r^{\prime}$, respectively, (i.e. $k_r^{} =
\kappa_r^{} + \kappa_r^{\prime}$). The observable under study is
usually the normalized fluorescence decay kinetics (FDK) $I_{_{{\rm
S}_1}}\!(t)/I_{_{{\rm S}_1}}\!(0)$ from ${\rm S}_1^{}$-state,
determined by the ${\rm S}_1^{}$-state population $p_{s}^{}(t)$ [for
which $p_{s}^{}(0) = 1$]: $I_{_{{\rm S}_1}}\! (t) =
\kappa_{r}p_{s}^{}(t)$, so that ${\bar I}_{_{{\rm S}_1}}\!(t) =
I_{_{{\rm S}_1}}\!(t)/I_{_{{\rm S}_1}}\!(0) = p_{s}^{}(t)$. In
accordance with this formula the characteristic features of singlet
fission kinetics are analyzed by comparison of the experimental FDK
${\bar I}_{_{{\rm S}_1}}\!(t)$ with the theoretically calculated
dependence $p_{s}^{}(t)$.

Detailed theoretical investigation of fission kinetics is performed
in a large number of papers \cite{Mi1,Sw1,Sw2,Mer1,Su}. Note,
however, that significant part of theoretical works concern the
analysis of the first stage of the process and, in particular, the
accurate evaluation of the rate of singlet splitting $({{\rm
S}_{0}^{}+{\rm S}_{_{\!}1\!}^{^{\!{*}}}}) \stackrel{k_{\!-\!s}^{}}
{\rightarrow}{\rm [TT]}$ \cite{Mi1,Sw1}. As for later stages
(essentially controlled by geminate spin/space evolution of TT-pair)
they are studied thoroughly as well, though results of the studies
are typically represented in fairly complicated mathematical form
not quite suitable for describing experiments. For this reason the
majority of experimental results are treated within the
above-discussed simplified model (\ref{form0}) \cite{Sw2,Mer1,Bou}.

Recent investigations \cite{Bar1,Bar2,Bar3,Bar4} show, however, that
the simplified model (of first order processes) is not able to
properly describe important specific features of the FDK ${\bar
I}_{_{{\rm S}_1}}\!(t)$, observed in some organic semiconductors, in
particular, the long time behavior of the FDK, which is found to be
close to the inverse-power type one.

In this work the generalized model is proposed, which allows for
accurate description of the effect of three-dimensional diffusive
migration of T-excitons (in [T+T]-state) on the FDK. The model is
shown to significantly improve the agreement between theoretical and
experimental FDK, especially at long times. Good accuracy and
potentialities of this model are demonstrated by analyzing the
above-mentioned FDK ${\bar I}_{_{{\rm S}_1}}\!(t)$, measured for a
number of organic semiconductors \cite{Bar1,Bar2,Bar3,Bar4}.

\section{Model of singlet fission}

To describe the important specific features of the FDK we propose
the generalized model of singlet fission (\ref{form0}). In the model
the first stage ${{\rm S}_{0}^{}+{\rm S}_{_{\!}1\!}^{^{\!{*}}}}
\rightleftarrows {\rm [TT]}$ is treated as a conventional first
order process. As to the second stage ${\rm [TT]} \rightleftarrows
{\rm [T+T]}$, it is described within the two-state approach
\cite{Shu1,Shu2,Shu3}, developed earlier to analyze the diffusive
escape of a particle from a potential well (intermediate state). In
this approach the spatial evolution of geminate TT-pairs is treated
as transitions between two states: (intermediate) [TT]-state of
coupled T-excitons and [T+T]-state of separated T-excitons,
undergoing isotropic three-dimensional diffusive migration.

The population $p_{s}^{}(t)$ of ${\rm
S}_{_{\!}1\!}^{^{\!{*}}}$-state is controlled by the spin/space
evolution of TT-pairs in [TT]- and [T+T]-states, described by the
spin matrix $\sigma (t)$ and spin density matrix $\rho (r,t)$,
respectively (here $r$ is the interexciton distance). These matrices
satisfy the stochastic Liouville equation \cite{St}, which in the
two-state approach is written in the form of three coupled equations
\begin{subequations}
\begin{eqnarray}
\dot p_{s}^{} &=& - (k_{\!-s}\! + k_r{})p_{s}^{} +
{\rm Tr}[{ {\cal P}}_{s}^{}
({\hat {\cal K}}_{s}^{}\sigma)] \label{form2a}\\
\dot \sigma &=& {\cal S}_l^{-1}{\cal K}_{_{\!+}\!} {\rho}_{l}^{}
- ({\hat {\cal L}}_{c}^{}\! +
_{\!}{\hat {K}}_{s}^{} + {\cal K}_{_{\!-}}\!)\sigma +
k_{\!-s}{{\cal P}}_{s}^{} p_{s}^{}\label{form2b} \qquad\\
\dot \rho &=& ({\cal D} \nabla_{\!r}^{2\!} -
_{\!}{\hat{\cal L}}_e^{}) \rho +
({\cal S}_l^{} {\cal K}_{_{\!-}\!}\sigma -
_{\!}{K}_{_{\!+\!}}\rho_{l}^{})
\delta (r_{\!}-_{\!}l)\label{form2c}\qquad
\end{eqnarray}
\end{subequations}
where $\:{\cal S}_l = (4\pi l_{}^{2})_{}^{-1}$, ${\cal D}$ is the
coefficient of relative diffusion of excitons, $\nabla_{\!r}^2 =
r^{-2}\partial_{r} (r^{2}\partial_{r})$ is the radial part of the
Laplace operator, and $\rho_{l}^{}(t) = \rho (l,t)$.

The terms proportional to rates ${\cal K}_{_{\!\pm}}$ represent the
above-mentioned transitions between the intermediate [TT]-state
($c$-state) and [T+T]-state ($e$-state) of freely diffusing
T-excitons. Values of ${\cal K}_{_{\!\pm}}$ satisfy the detail
balance relation \cite{Shu3}. In our work we will consider the
realistic limit of fast spatial relaxation of TT-pairs in
[TT]-state, in which this relation is represented as \cite{Shu3}
\begin{equation} \label{form3}
{\cal K}_{_{\!\pm}} \rightarrow \infty \quad \mbox{and}
\quad {\cal K}_{_{\!+}}/{\cal K}_{_{\!-}} =
\lambda_{e} = Z_w^{}/l_{}^2,
\end{equation}
where $Z_w = \int_{r \sim l}dr \, r^2 e^{-U(r)/(k_B^{} T)}$ is the
partition function for [TT]-state [in the well $U(r)$] and $l$ is
the radius of the state. The parameter $\lambda_{e}$ controls
effects of [TT]- and [T+T]-states on the FDK: for significant
TT-coupling in [TT]-state, when $\xi_{e} = \sqrt{l/\lambda_{e}} < 1$
[eq. (\ref{res5})], strong [TT]-effect and weak [T+T]-effect
(T-migration effect) is predicted. In the case of weak coupling
($\xi_{e} > 1$), on the contrary, small [TT]- and marked
[T+T]-effects are expected (see Sec. 4.1).


In eqs (\ref{form2a})-(\ref{form2c}) the terms
\begin{equation} \label{form4}
k_{\!-s}{{\cal P}}_{s}^{} p_{s}^{} \;\;\mbox{and}\;\;
\hat {\cal K}_{s}^{}\sigma = (1/2)k_{s}{}
({\cal P}_{\!s}^{}\sigma + \sigma {\cal P}_{\!s}^{}),
\end{equation}
with the projection operator ${\cal P}_{\!s}^{} = |S
\rangle\!\langle S|$ onto the singlet ($S$) state of TT-pair,
describe the spin-selective TT-pair generation (resulting from the
decay of ${{\rm S}_{_{\!}1\!}^{^{\!{*}}}}$-state with the rate
$k_{\!-\!s}^{}$) and annihilation (with the rate $k_{\!s}^{}$)
\cite{Sw2,St}, respectively. Operators $\hat {\cal L}_{c}^{}$ and
$\hat {\cal L}_{e}^{}$, defined as ($\hbar = 1$)
\begin{equation} \label{form5}
\hat {\cal L}_{\nu}^{}\rho = \hat {\cal W}_{\nu}^{}\rho +
i(H_{\nu}^{}\rho - \rho H_{\nu}^{}), \quad (\nu = c, e),
\end{equation}
for any spin matrix $\rho$, control the spin evolution in [TT]- and
[T+T]-states, respectively. In formula (\ref{form5}) $\hat {\cal
W}_{\nu}^{}$ is the operator of spin relaxation in $\nu$-state of
TT-pair (its explicit form is specified below), and
\begin{equation} \label{form6}
H_{\nu}^{} = g\beta B (S_{a\!}^{z} + S_{b\!}^{z})
+ H_{T_\nu}^{a} + H_{T_\nu}^{b},
\end{equation}
is the spin Hamiltonian of the TT-pair in the magnetic field ${\bf
B}$ (directed along the axis $z$), in which the first term describes
the Zeeman interaction of spins with the field ${\bf B}$ and
$H_{T_\nu}^{\mu}$ is the zero-field-splitting interaction (ZFSI)  in
exciton $\mu, \, (\mu = a, b)$, in the state $\nu, \, (\nu = c, e)$.

For example, in $c$-state $H_{T_c}^{\mu} \equiv H_{T}^{\mu}, \: (\mu
= a, b),$ with
\begin{equation} \label{form7}
H_{T}^{\mu} = D_{\mu}^{}[(S_{\mu\!}^{z_\mu\!})_{}^{2}\!
-\!S_{\mu}^{2}/3] + E_{\mu}^{}[(S_{\mu\!}^{x_\mu\!})_{}^{2}\! -
\!(S_{\mu\!}^{y_\mu\!})_{}^{2}].
\end{equation}
Here $S_{\mu}^{j_\mu}$ is the projection of the spin of the exciton
$\mu$ along the eigenaxis $j_\mu,\, (j_\mu = x_\mu, y_\mu, z_\mu)$,
of the ZFSI-tensor of T-exciton in $c$-state \cite{Sw2,St}.

As for the ZFSI $H_{T_e}^{\mu}$ in $e$-state, its precise form
depends on the process under study. In our work we will consider two
most interesting and experimentally investigated
\cite{Sw2,Bar1,Bar2,Bar3,Bar4} types of them, for which the form of
$H_{T_e}^{\mu}$ can easily be obtained:

(1) Singlet fussion in amorphous solids, in which in $e$-state
T-excitons undergo fast hopping over chaotically oriented molecules,
resulting in the average of the ZFSI (i.e. $H_{T_e}^{\mu} = 0$), and
in fast spin relaxation (see below);

(2) Singlet fussion in (molecular) crystals, for which
$H_{T_c}^{\mu} = H_{T_e}^{\mu} = H_{T}^{}$.

The TT-spin evolution, governed by spin-Hamiltonians $H_{\nu}^{}$,
is described with the complete basis of 9 spin states, represented
as products $|j_a\!j_b \rangle = |j_a \rangle |j_b \rangle$ of those
$|j_{\mu} \rangle,\:(j_{\mu} = 1-3),\,$ for T-excitons. Hereafter it
is convenient to use the eigenstates of the Zeeman Hamiltonian
$|j_{\mu} = 0, \pm \rangle$ (defined as $S_{z}^{}|j_{\mu}\rangle =
j_{\mu} |j_{\mu}\rangle$) or those of the ZFSI (\ref{form7})
$|j_{\mu} = x_{\mu}^{}, y_{\mu}^{}, z_{\mu}^{}\rangle $ (defined by
$S_{j_\mu}|j_{\mu} \rangle = 0 $) \cite{Sw2}. In particular, within
these two bases $|S\rangle$-state of TT-pair is represented as
\cite{Sw2}
\begin{equation} \label{form8}
|S\rangle_{\!} =_{\!} \frac{1}{\sqrt{3}}\!
(|0 0\rangle\! -\!|\!+\! -\!\rangle\!
 - \!|\!-\! +\!\rangle)\!
= \!\frac{1}{\sqrt{3}}\!\sum_{j=x,y,z}\!\! |jj\rangle.
\end{equation}

The kinetic scheme (\ref{form0}) implies the initial condition for
eqs (\ref{form2a})-(\ref{form2c})
\begin{equation} \label{form9}
p_{s}^{} (t=0) = 1\;\; \mbox{and} \;\; \sigma
(t=0)=\rho(t=0)=0.
\end{equation}
These equations should be solved with the reflective boundary
condition for $\rho (r,t)$ at $r=l$: $\partial_{r}^{}\rho|_{r=l}^{}
= 0$.

The solution can be obtained by the Laplace transformation in time,
defined for any function $\varphi (t)$ as $\widetilde{\varphi}
(\epsilon) = \int_{0}^{\infty}\! dt \, \varphi (t)e^{-\epsilon t}$
and ${\varphi} (t) = (2\pi i)_{}^{-1} \int_{\!-i\infty}^{i\infty}\!
d\epsilon \, \widetilde{\varphi} (\epsilon)e^{\epsilon
t}$\cite{Shu2,Shu3} In particular, for $\widetilde {p}_{s}^{}
(\epsilon)$ we get the expression
\begin{equation} \label{res2}
\widetilde {p}_{s}^{} (\epsilon) =
\big\{\epsilon + k_{r\!s}^{} - k_{\!-s\!}^{} {\rm Tr}[
{\cal P}_{\!s}^{} \hat {\cal K}_{s}^{}
\hat {\cal G} (\epsilon) {\cal P}_{\!s}^{}]\big\}_{}^{-1}.
\end{equation}
Here ${\cal P}_{\!s}^{}$ is the projection operator, $k_{r\!s}^{} =
k_{r}^{} + k_{\!-s\!}^{\!}$ with $\,k_{r}^{} =
\kappa_{r}^{}+\kappa_{r}^{\prime}$, and
\begin{equation} \label{res3}
\hat {\cal G} (\epsilon) = [\epsilon + \hat {\cal L}_{c} +
\hat {\cal K}_{s}^{} + \hat {\cal K}_{e}^{}(\epsilon)]_{}^{-1}
\end{equation}
is the Laplace transform of the evolution function of [TT]-state,
decaying with the effective escape rate
\begin{equation} \label{res4}
\hat {\cal K}_{e}^{}(\epsilon) = k_{e}^{} +
\hat \kappa_{\epsilon}^{}  
\;\;\mbox{with} \;\;\hat \kappa_{\epsilon}^{} = 
\xi_e^{}[k_{e}^{}(\epsilon \!+\!\hat {\cal L}_e^{})]_{}^{1/2}\!,
\end{equation}
in which
\begin{equation} \label{res5}
k_{e}^{} = {\cal D} l/Z_w^{} \;\; \mbox{and} \;\;
\xi_e^{} = \sqrt{l_{}^{2}k_{e}^{}/{\cal D}}.
\end{equation}

It is worth noting that the rate $\hat {\cal K}_{e}^{}(\epsilon)$
essentially determines the kinetics of diffusion assisted
TT-annihilation \cite{Shu1,Shu2,Shu3} and, thus, the FDK. Of special
importance is the second $\epsilon$-dependent (non-analytic) term
$\hat \kappa_{\epsilon}^{}$, which is responsible for the
non-exponential behavior of the annihilation kinetics at
intermediate and long times. Of course,
specific features of this behavior are affected by spin evolution of
TT-pair, though mainly at times smaller than spin relaxation times.
Below we will analyze spin effects in two important limits of fast
and slow spin relaxation (Secs. IV.A. and IV.B.). Here we only note
that $\hat \kappa_{\epsilon}^{}$ increases with the increase of
$\xi_e^{}$ [eq. (\ref{res4})], and for $\xi_e^{} > 1$ the
T-migration effect on the FDK is expected to be very strong (see
below).

Calculation of $\widetilde {p}_{s}^{} (\epsilon)$ with eq.
(\ref{res2}) is a fairly complicated problem which, however, can be
simplified in the Johnson-Merrifield approximation (JMA)
\cite{Sw2,Mer1}.

\section{Johnson-Merrifield approximation}
The JMA allows one to reduce cumbersome operations with elements of
TT-spin density matrices to those with state populations only, i.e.
diagonal matrix elements in the basis of eigenstates of the
Hamiltonian $H_{c}^{}$ (\ref{form6}) (as discussed above). The weak
effect of non-diagonal elements (for $\| H_{c}^{}\|/k_{s,e}^{} \gg
1$ \cite{Shu4}) results from their fast oscillations.

To present JMA-results conveniently we introduce the additional
notation. Recall, that for any spin system with $N$ states
($|j_a\!j_b\rangle = |j_a\rangle |j_b \rangle$) eqs (\ref{form2b})
and (\ref{form2c}) are systems of $N_{}^2$ coupled equations for
elements of density matrices $\sigma$ and $\rho$, i.e. components of
vectors in the basis of "states" $|j_a\!j_b \rangle\langle
j'_a\!j'_b|$ in the Liouville space \cite{Blum}. In the JMA these
systems reduce to those of $N$ equations for components "along"
population eigenvectors
\begin{equation} \label{res6a}
\| j_a\!j_b \rangle\!\rangle \equiv |j_a\!j_b
\rangle\langle j_a\!j_b|,
\end{equation}
corresponding to diagonal elements of density matrices.

With this notation, the JMA-formula for $\widetilde{p}_{s}^{}$ is
given by
\begin{equation} \label{res6}
\widetilde{p}_{s}^{} (\epsilon)\! = \big[\epsilon + k_{r\!s}^{} -
N_{} k_{\!-\!s}^{}\langle\!\langle e\|{\hat K}_{\!s}^{}
\hat G (\epsilon){\hat P}_{\!s}^{}\|e\rangle\!\rangle\big]_{}^{-1}.
\end{equation}
Here $k_{r\!s}^{} = k_{r}^{} + k_{\!-s}^{}$, $N$ is the number of
spin states of TT-pair ($N = 9$), and
\begin{equation} \label{res6b}
\|e\rangle\!\rangle = N^{-1}\!\sum\nolimits_{j_{a}\!j_{b}}\!\!
\| {j_{a}\!j_{b}} \rangle\!\rangle
\;\:\mbox{and}\;\: \langle\!\langle e \| =
\sum\nolimits_{j_a\! j_b}\!\!\langle\!\langle {j_{a}\!j_{b}}\|
\end{equation}
are the normalized equilibrium state vector and the corresponding
adjoined one $( \langle\!\langle e \|e\rangle\!\rangle =1)$, and
\begin{equation} \label{res8}
\hat K_{s}^{} = k_s^{}\hat P_{s}^{}, \:\;\mbox{with}\:\;
{\hat P}_{\!s\,}^{} = \sum\nolimits_{\!j_{a\!},j_b}\!
\!C_{\!j_a\!j_b}^{S}\| j_a\!j_b \rangle\!\rangle \langle\!\langle j_a
\!j_b\|,
\end{equation}
is the annihilation rate matrix, proportional to the matrix ${\hat
P}_{\!s\,}^{}$ of ($B$-dependent) weights $C_{\!j_a\!j_b}^{S} =
|\langle S|j_a^{}\!j_b^{}\rangle|_{}^2$ of $S$-state in states
$|j_a^{}\!j_b^{}\rangle$ of the TT-pair (satisfying the
normalization condition $\sum_{j_a\!j_b}\!\! C_{j_a\!j_b}^{S} = 1$),
\begin{equation} \label{res9}
\hat G (\epsilon) = \Big\{\epsilon + k_{e\!}^{\!} + \hat K_{s\!}^{\!}
+ \hat {W}_{\!c\!}^{\!} +
\xi_e^{}\big[k_{e}^{} (\epsilon\! + \!\hat W_{\!e}^{})\big]_{}^{1/2}
\!\Big\}_{\!}^{\!-1}
\end{equation}
is the evolution function for combined ($ce$)-state, in which $\hat
W_{\!\nu}^{}$ are matrices of spin-lattice relaxation for TT-pairs
in states $\nu = c, e$. In our further analysis we well use matrices
$\hat W_{\!\nu}^{}$ in a simplified analytical form:
\begin{equation} \label{res10}
\hat W_{\!\nu}^{} = w_{\nu\!}^{\!} (B)\hat Q \;\;
\mbox{with}\;\; \hat Q = \hat E -
\|e\rangle\!\rangle_{\!} \langle\!\langle e\|. 
\end{equation}
In this expression $\hat Q$ is the projection operator ($\hat Q_{}^2
= \hat Q$) with $\hat E = \sum\nolimits_{j_{a\!},_{\!}j_b}\!\|
j_a\!j_b \rangle\!\rangle_{\!}\langle\!\langle j_a \!j_b\|$. The
form (\ref{res10}) is sufficient for semiquantitative treatment of
the FDK (see below). Noteworthy ia that for small $\epsilon <
w_c^{}$, describing the FDK at long times $t < w_c^{-1}$, the effect
of the relaxation (\ref{res10}) is described by the effective
relaxation operator $\hat W_{\!r}^{} = \hat {W}_{\!c}^{} +
\xi_e^{}\big(k_{e}^{} \hat W_{\!e}^{}\big)_{}^{\!1/2}\!,$ also
represented in the form (\ref{res10})
\begin{equation} \label{res10a}
\hat W_{\!r}^{} = w_r^{} \hat Q, \;\;\mbox{where}\;\; w_r^{} =  w_c^{} +
\xi_e^{} \sqrt{k_e w_e^{}}.
\end{equation}

In the model (\ref{res10}) formula for $\widetilde p_{s}^{}
(\epsilon)$ is very complicated in general. In some systems,
however, the FDK at $B = 0$ and $B \gg B_s^{} =
\|H_{T}^{\mu}\|/(g\beta)$ can conveniently be analyzed within the
simple model of $n_r$ equally reactive states, denoted as $\|
j_{r}^{} \rangle\!\rangle$, which turn out to be equivalent and
equally contributing to ${p}_{s}^{} (t)$. Similarly equivalent are
also $n_n = N - n_r^{}$ nonreactive states $\| j_{n}^{}
\rangle\!\rangle$. These equivalences enable one to reduce the
problem of $N$ coupled states to that of the pair of states
\begin{equation} \label{two1}
\|e_{\alpha}^{}\rangle\!\rangle =
n_{\alpha}^{\!-\!1}\!\sum\nolimits_{\!j_{\alpha}^{}=1}^{n_{\alpha}}\!
\| {j}_{\alpha}^{} \rangle\!\rangle,\;\,
\langle\!\langle e_{\alpha}^{}\| =
\!\sum\nolimits_{\!j_{\alpha}^{}=1}^{n_{\alpha}}\!
\langle\!\langle {j}_{\alpha}^{} \| ,
\end{equation}
where $\alpha = r,n$.

In the basis (\ref{two1}) matrices $\hat W_{\!\nu}^{}$ and $\hat
K_{s}^{}$ are written as
\begin{eqnarray} 
\hat W_{\!\nu}^{}\,  &=& 
w_{\nu}^{}
\big[\zeta_{r}^{}({\hat P}_{\!rr}^{}\!
-{\hat P}_{\!nr}^{})\!
+ \zeta_{n}^{}({\hat P}_{\!nn}^{}\!
-{\hat P}_{\!rn}^{})\big], \label{two2a}\\
\hat K_{\!s}^{}\, &=& k_{s}^{}{\hat P}_{\!rr}^{}, \;\;\mbox{with}\;\;\:
{\hat P}_{\!\alpha\alpha'}^{}  =  
\|e_{\alpha}^{}\rangle\!\rangle \langle\!\langle e_{\alpha'}^{}\!\|
\quad \label{two2b}
\end{eqnarray}
and $\zeta_{\alpha}^{} = 
1-n_{\alpha}/N,\:(\alpha = r, n)$.

Within this approach of two effective states general formula
(\ref{res6}) reduces to a fairly simple analytical one:
\begin{equation} \label{two3}
\widetilde{p}_{s}^{} (\epsilon)\! =
\bigg[\epsilon + k_{r\!s\!}^{\!} -
\frac{(k_{\!-\!s}^{}\kappa_{\!s}^{})g_{r}^{}(\epsilon)}{1\! - \!
\zeta_{r}^{}\zeta_{n}^{}\kappa_{n\!}^{2\!}(\epsilon)
g_{r\!}^{\!}(\epsilon)g_{n\!}^{\!}(\epsilon)}\bigg]_{}^{-1}\!\!,
\end{equation}
where $k_{r\!s}^{} = k_{r\!}^{\!} + k_{\!-\!s}^{}$,
$\:\kappa_{s\!}^{\!} = k_{s}^{}/n_{r}^{}$, and
\begin{eqnarray}
\kappa_{n\!}^{\!}(\epsilon) &=& w_{c\!}^{\!} + \xi_e^{}
\big(\sqrt{k_e^{}(\epsilon + w_e^{})}
- \sqrt{k_e^{}\epsilon}\,\big)\,,\label{two4a} \\
g_{\alpha}^{}(\epsilon) &=&
\big[\epsilon + \kappa_{\!s_\alpha\!}^{\!} + k_{e\!}^{\!} +
\kappa_{n\!}^{\!}\zeta_{\alpha}^{} +
\xi_e^{}\!\sqrt{k_{e}^{}\epsilon}\,\big]_{}^{-1}\!,\qquad \label{two4b}
\end{eqnarray}
with $\alpha = r, n;\,$  and  $\,\kappa_{s_{r}\!}^{\!} =
\kappa_{s\!}^{\!},\:\kappa_{\!s_{n}\!}^{\!} = 0$.

Noteworthy is that any additional relaxation within sets of reactive
($\| j_{r}^{} \rangle\!\rangle$) and/or non-reactive ($\| j_{n}^{}
\rangle\!\rangle$) states, separately, does not affect $p_{s}^{}
(t)$ because of equipopulation of states in these sets. This means
that the FDK $p_{s}^{} (t)$ is essentially determined by relaxation
transitions between states of different sets, rates of which can be
different for small and large $B$ ($w_e^{_0}$ and $\bar w_e^{}$)
(Sec. IV.A).

Formulas (\ref{res6})-(\ref{two4b}) are suitable for analyzing the
T-exciton migration effect on the FDK. The analysis requires
specification of parameters of the model (\ref{two1})-(\ref{two4b}),
which are essentially determined by the magnetic field $B$. In our
further analysis, for brevity, we will discuss the FDK for $B = 0$
and $B \gg B_{s}^{} = \|H_{T}^{\mu}\|/(g\beta)$ only.

\section{Results and Discussion}
\subsection{FDK in amorphous organic semiconductors}
In the proposed model (\ref{res6}) specific features of the FDK in
amorphous semiconductors result, essentially, from strong disorder
of molecule orientations, i.e. orientations of ZFSI eigenaxes.

{\it 1. FDK in the absence of magnetic field $(B = 0)$.}\, In the
JMA the orientational distribution of excitons manifests itself in
the spread of TT-annihilation rates $K_{s_{j_a \!j_b}} =
\langle\!\langle j_a \!j_b\|\hat K_s^{} \| j_a\!j_b \rangle\!\rangle
= k_s^{} C_{\!j_a\!j_b}^{S}$ in $c$-state ([TT]) with the same
distribution function for all states $\| j_a\!j_b \rangle\!\rangle,
(j = x, y, z)$, i.e. the same mean value $\langle K_{s}^{}\rangle =
\langle K_{s_{j_a \!j_b}}\rangle = k_s^{}\langle
{C}_{\!j_a\!j_b}^{S}\rangle$ (where $ \langle
{C}_{\!j_a\!j_b}^{S}\rangle= (1/3)\langle
\cos^{2\!}\theta_{\!j_a\!j_b\!} \rangle_{\theta_{j_aj_b}} = 1/9$ is
the average over the angle $\theta_{\!j_a\!j_b}$ between axes $j_a$
and $j_b$) and dispersion $\Delta_{^K}^{} = \sqrt{\langle
K_{s}^{2}\rangle - \langle K_{s}^{}\rangle_{}^{2}} \approx
0.1\,k_s^{}$.

In general, the spread of rates $K_{s_{j_a \!j_b}}$ significantly
complicates the evaluation of the FDK. Fortunately, in many
amorphous semiconductors spin relaxation rates $w_{\nu}^{}(B=0) =
w_{\nu}^{_{0}}$, ($\nu = c,e$), are expected to be fairly large
(Sec. IV.A.2), resulting in large $w_r^{}$ [see eq. (\ref{res10a})]:
\begin{equation} \label{amo1}
w_r^{}(B=0) = w_{r}^{_0}
\gg \Delta_{^K}^{}.
\end{equation}
So fast relaxation leads to efficient averaging the rates $K_{s_{j_a
\!j_b}}$, i.e high accuracy of the approximation $\hat K_{s}^{}
\approx \kappa_{\!s}^{_{0}}\hat E$, where $\kappa_{\!s}^{_{0}} =
\langle\!\langle e \|\hat K_{\!s}^{} \|e\rangle\!\rangle = \langle
K_{s}^{}\rangle = k_s^{}/9$. In such a case [corresponding to
$n_{r}^{} = N = 9$, $\zeta_r^{} = 0,\,\zeta_n^{} = 1,$ and
$\kappa_s^{} = \kappa_{\!s}^{_0}$ in eq. (\ref{two3})]
$\widetilde{p}_{s}^{} (\epsilon)$ is given by
\begin{equation} \label{amo2}
\widetilde{p}_{s}^{} (\epsilon)\! =
\bigg(\epsilon + k_{r\!s\!}^{\!} - \frac{k_{\!-\!s}^{}\kappa_{\!s}^{_0}}
{\epsilon + \kappa_{\!s}^{_0} + k_{e}^{} +
\xi_e^{} \!\sqrt{k_{e}^{}\epsilon}}\bigg)_{}^{\!\!-1}
\end{equation}
with $k_{r\!s}^{} = k_{r\!}^{\!} + k_{\!-\!s}^{}$ and
$\;\kappa_{\!s}^{_{0}} = k_{s}^{}/9.$

The inverse Laplace transformation of $\widetilde{p}_{s}^{}
(\epsilon)$ (\ref{amo2}) yields the FDK ${p}_{s}^{}(t)$, predicting
the T-exciton migration effect, which shows itself, in particular,
in the slow long time dependence: ${p}_{s}^{}(t) \sim
\xi_e^{}t^{-3/2}$, resulted from the nonanalytic behavior of
$\widetilde{p}_{s}^{} (\epsilon)$ at small $\epsilon$:
$\widetilde{p}_{s}^{} (0) - \widetilde{p}_{s}^{} (\epsilon) \sim
\xi_e^{}\sqrt{\epsilon}$ \cite{Shu2,Shu3}. The amplitude of the
migration effect is essentially determined by the value of the
parameter $\xi_e^{}$, as stated in Sec. II, so that for large
$\xi_e^{} \gtrsim 1$ the effect is fairly strong and clearly
distinguishable even at relatively short times \cite{Shu2,Shu3}.

\begin{figure}[h]
\setlength{\unitlength}{1cm} 
\includegraphics[height=6.1 cm,width=7.2cm]{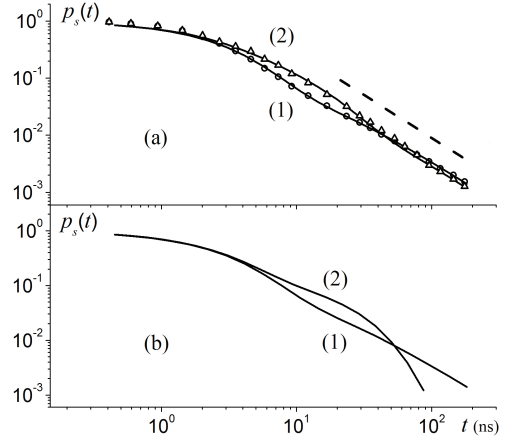}
\caption{(a) Comparison of the experimental FDK for amorphous
rubrene films\cite{Bar1} at $B = 0$ (circles) and $B = 8\; {\rm kG}$
(triangles) with the FDK $p_s^{}(t)$, calculated by eq. (\ref{amo2})
for $B = 0$ (line 1) and eq. (\ref{two3}) for $B \gg B_s^{}$ (line 2),
using the rate value $k_{r\!s\!}^{\!} = 0.37\; {\rm ns}_{}^{-1}$, the vector
of other parameters ${\bf z} = (0.2; 1.44; 0. 14; 1.7)$ [eq.
(\ref{amo5})], and relaxation rates $\bar w_c^{}/k_{r\!s\!}^{\!} = 0.04;\:
\bar w_e^{}/k_{r\!s\!}^{\!} = 0.5 $. The dependence $p_s^{}(t) =
A t_{}^{-3/2}$, with arbitrary constant $A$, is also displayed
(dashed line). (b) Comparison of the FDK $p_s^{}(t)$
for $B = 0$ (line 1 as in Fig. 1a) with the conventional
variant of the FDK (determined by first order processes),
evaluated with eq. (\ref{amo2}) for the same parameters, except
$\xi_e^{} = 0$, i.e. excluding migration effect (line 2).}
\end{figure}

High accuracy of formula (\ref{amo2}) in the limit (\ref{amo1}) is
demonstrated in Fig. 1a by comparison of the theoretical dependence
${p}_{s}^{}(t)$ with the normalized FDK ${\bar I}_{_{{\rm
S}_1}}\!(t) = I_{_{{\rm S}_1}}\!(t)/I_{_{{\rm S}_1}}\!(0)$ (Sec. I),
measured for amorphous rubrene \cite{Bar1}, in which the
spin-lattice relaxation is expected to be very fast (see Sec.
IV.A.2). The function ${p}_{s}^{}(t)$ is calculated with eq.
(\ref{amo2}) by adjusting the rate $k_{r\!s\!}^{\!}$ and other
rates, represented as a set of dimensionless parameters
\begin{equation} \label{amo5}
{\bf z} = (z_{r}^{}, z_{s}^{}, z_{e}^{}, \xi_{e}^{}),\;\;
\mbox{where}\;\; z_{q}^{} = k_{q}^{}/k_{r\!s}^{}
\end{equation}
with $q = r, s, e$. Good agreement is found at all studied times,
including long times when ${p}_{s}^{}(t) \sim t_{}^{-3/2}$. The FDK
is also calculated in the conventional model of first order
processes (\ref{amo2}), i.e. a particular variant of the general one
with $\xi_{e}^{} = 0$, for the same values of parameters $z_{q}^{}$.
This model is seen to predict too sharp decrease of ${p}_{s}^{}(t)$
at long times (Fig. 1b), which is, clearly, valid for any values of
parameters of the model.

{\it 2. FDK for strong magnetic fields.}\, In the limit of strong
field $B \gg B_s^{} = \| H_{T}^{\mu}\|/(g\beta)$ there are 3 spin
states with singlet character: $|1\rangle \approx
|0_a^{}0_b^{}\rangle$, $|2\rangle \approx
|\!\!+_{\!a\!\!}^{\!\!}\!-_{\!b}^{}\rangle$, and $|3\rangle \approx
|\!\!-_{\!a\!\!}^{\!\!}\!+_b^{}\rangle$. Note that for two
independently oriented T-excitons at large, but finite, $B$ the
states $|2\rangle$ and $|3\rangle$  are, in general, non-degenerate
with small splitting $\omega_{23}^{} \sim D[D/(g\beta B)] \ll D$,
where $D=D_a=D_b$ is the ZFSI-parameter of T-excitons. In this
estimation it is taken into account that for separated aromatic
molecules (typically existing in amorphous solids) $D \gg E$
\cite{Sw2}. In the presence of this nearly degenerate pair of states
the TT-spin evolution is, nevertheless, described by the JMA, if
$\omega_{23}^{}/k_e^{} \sim \omega_{23}^{}/\bar\kappa_s^{} > 1$
\cite{Shu4}, where $\bar\kappa_s^{} = k_s^{}/3$. For studied
amorphous rubrene films, in which $\bar \kappa_s^{} \approx
3\,k_e^{} \approx 0.15\, {\rm ns}^{-1}$ (see Fig. 1a) and $D \approx
0.6\, {\rm kG}$ \cite{Shu01}, this criterion predicts the validity
of the JMA at $B < 40\, {\rm kG}$. Hence in the experimentally
studied case \cite{Bar1} $B = \bar B = 8.1\, {\rm kG}$ the JMA is
valid, and the FDK is described by the model
(\ref{two3})-(\ref{two4b}) with $n_r = 3,\;n_n = 6, \;$ and
$\;\kappa_s^{} = \bar \kappa_s^{} = k_s^{}/n_r = k_s^{}/3$ [\,for
$C_{\!jj}^{S} \approx 1/3,\; (j = 1-3)$\,].

Figure 1a displays ${p}_{s}^{}(t)$, calculated for $B \gg B_s^{}$ by
Eqs. (\ref{two3})-(\ref{two4b}) with the same values of parameters
(\ref{amo5}) as those applied above for $B = 0$. The FDK
${p}_{s}^{}(t)$ agrees fairly well with the experimental one at $B =
\bar B = 8.1\, {\rm kG}$ \cite{Bar1}. The agreement is obtained,
assuming fast spin relaxation with rates $w_e(B = \bar B) = \bar
w_{e}^{} \approx \,0.185\; {\rm ns}_{}^{-1}$ and $w_c(B = \bar B) =
\bar w_{c}^{} \approx \,0.015\; {\rm ns}_{}^{-1}$, induced by
fluctuating ZFSI in T-excitons undergoing stochastic migration
\cite{Shu5}.

For so large rates $\bar w_{c,e}^{}$ the effective spin relaxation
rate (\ref{res10a}) $\bar w_r^{} = w_r^{}(B = \bar B)$ is also
fairly large: $\bar w_r^{} \approx 3.4\,\Delta_{^K}^{}$. This
relation is the important argument in favor of good accuracy of the
averaging approximation $K_{s_{\!j_a\!j_b}} \approx \langle
K_{s_{j_a \!j_b}}\rangle = \kappa_s^{_0}$ (applied above for $B =
0$), especially taking into account that $w_{r}^{_0} \gg \bar
w_{r}^{}$ [because of expected inequalities $w_{\nu}^{_0}/\bar
w_{\nu}^{} \gg 1$ ($\nu = c, e$)] and therefore $w_r^{_0} \gg
3.4\,\Delta_{^K}^{}$. As to the estimation $w_{\nu}^{_0}/\bar
w_{\nu}^{} \gg 1$, for the system under study in the strong magnetic
field $\bar B = 8.1\, {\rm kG}$ it can be obtained using the
relation $w_r^{_0}/\bar w_r^{} \sim (g\beta \bar B)_{}^2\tau_{c}^2$
(where $\tau_c^{}$ is the correlation time of ZFSI-fluctuations),
taking into account that realistic values $\tau_c^{} \gtrsim
10^{-1}\, {\rm ns}$ \cite{Sw2}.

Noteworthy is also that for typical case of small deactivation rates
$k_r^{} \ll k_{\!-s}^{}$ at long times ($t
> 1/k_{\!s}^{},1/\kappa_{s}^{_0}$) the reversible initial splitting
of the singlet state, $({\rm S}_{0}^{}+{\rm
S}_{_{\!}1\!}^{^{\!*}})^{^{\!{}}}\,
\rightleftarrows_{\!\!\!\!\!\!_{{\hat K_{\!s}}}}
^{\!\!\!\!\!\!\!\!^{^{{k_{-\!s}}}}}\, [{\rm TT}]\,$ [see eq.
(\ref{form0})], results in the effective reduction of annihilation
rate $\bar \kappa_{s}^{_0} \sim \kappa_{s}^{_0}
(k_{r}^{}/k_{\!-s}^{}) \ll \kappa_{s}^{_0}$. For these values of
rates $\bar \kappa_{s}^{_0}$ the rate averaging condition $w_r^{_0}
\gg \,\Delta_{^K}^{}$ can easily be satisfied, in reality.

\subsection{FDK in molecular crystals}

The proposed approach is quite suitable for evaluating the FDK in
molecular crystals as well. The calculation of ${p}_{s}^{}(t)$ for
molecular crystals is somewhat simpler than for amorphous
semiconductors because of slow spin relaxation in T-excitons in
crystals \cite{Sa}: $w_{c,e}^{} \ll k_{q}^{},\:(q = -s, s, e)$.
According to theoretical and experimental estimations
\cite{Sa,Wol,Con} in molecular crystals $w_{c,e} \lesssim 10^{7}_{}
\; {\rm s}^{-1}_{}$, which means, for example, that  in experiments
under study \cite{Bar4} the effect of spin relaxation on the FDK in
crystals can be neglected at times $t \lesssim 10_{}^{-8}\; {\rm
s}$.

Especially simple expressions for ${p}_{s}^{}(t)$ can be found in
the considered cases $B = 0$ and $B \gg B_s^{}$. For certainly we
will discuss homofission processes \cite{Sw2}, i.e. splitting into
two identical T-excitons.

{\it 1. FDK in the absence of magnetic field $(B = 0)$.}\, For $B =
0$ within the JMA ${p}_{s}^{}(t)$ is determined by spatial evolution
of T-excitons in three equally reactive (population) states only:
$\| xx \rangle\!\rangle,\, \| yy \rangle\!\rangle,\,$ and $\,\| zz
\rangle\!\rangle$. In such a case the FDK is described by the
universal expression (\ref{amo2}) with $\kappa_s^{} = k_s^{}/3$
(corresponding to $n_r^{} = N = 3,\;\; n_n = 0$).

\begin{figure}[h]
\setlength{\unitlength}{1cm} 
\includegraphics[height=5.9cm,width=6.7cm]{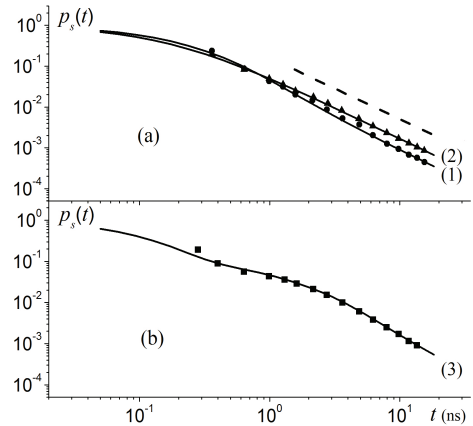}
\caption{Comparison of the experimental FDK, measured for
tetracene: (a) single crystal (circles), annealed PCF (triangles),
and (b) PCF film (squares) at $B = 0$,\cite{Bar4} with the FDK
$p_s^{}(t)$ (full lines $1, 2$ and $3$, respectively), calculated
using eq (\ref{amo2}) for parameters: (1) $k_{r\!s\!}^{\!} = 6.5\:
{\rm ns}_{}^{-1}$ and ${\bf z}_1^{} = (0.032; 2.9; 0.97; 1.7)$; (2)
$k_{r\!s\!}^{\!} = 8.5\: {\rm ns}_{}^{-1}$ and ${\bf z}_2^{} =
(0.028; 2.9; 0. 56; 2.2)$; and (3) $k_{r\!s\!}^{\!} = 10.0 \: {\rm
ns}_{}^{-1}$ and ${\bf z}_3^{} = (0.003; 0.4; 0.08; 0.65)$. The
dependence $p_s^{}(t) = At_{}^{-3/2}$ (with arbitrary amplitude $A$)
is also displayed in Fig. 2a for comparison (dashed line).}
\end{figure}

Formula (\ref{amo2}) allows for fairly accurate description of the
FDK, measured for tetracene: single crystal, polycrystalline film
(PCF), and annealed PCF  at $B = 0$ and times $t < 20 \; {\rm ns}$
(Figs 2a and 2b) \cite{Bar4}. In our work we will restrict ourselves
to pointing out most important features of the T-exciton migration
manifestation, concentrating on the behavior of the FDK at
relatively long times ($t > 1\; {\rm ns}$) influenced by migration
most strongly.

a) Shapes of the FDK ${\bar I}_{_{{\rm S}_1}}\!(t) = I_{_{{\rm
S}_1}}\!(t)/I_{_{{\rm S}_1}}\!(0)$ for single crystal and annealed
PCF are similar with close values of corresponding kinetic
parameters (\ref{amo5}) for both systems (Fig. 2a).

b) For the PCF the FDK-shape markedly differs from those for two
other systems (Fig. 2b). The difference results from fast decay of
${\rm S}_{1}^{}$-state ($k_{\!-\!s}^{} \gg k_{s}^{}, k_{e}^{}$) and
weaker effect of T-exciton migration in the PCF, showing itself in
the value $\xi_e^{} \approx 0.65$ smaller than those for other
studied systems ($\xi_e^{} \approx 1.5 - 2.0$).

c) In all three systems the long time tail of the FDK (at $t
> k_{s,e}^{-1}$) is fairly strongly affected by stochastic migration
of T-excitons. This effect manifests itself in the inverse-time
behavior of ${p}_{s}^{}(t)$: ${p}_{s}^{}(t) \sim t_{}^{-3/2}$.

d) Fitting of the experimental FDKs ${\bar I}_{_{{\rm S}_1}}\!(t)$
yields values $k_{rs}^{} \approx k_{\!-\!s}^{} \approx 6\, -\, 10\;
{\rm ns}_{}^{-1}$ for all three systems, close to those, measured at
short times $t < 0.8 \;{\rm ns}$ \cite{Bar4}.

{\it 2. FDK for strong magnetic fields.}\, At strong magnetic fields
$B \gg B_s^{}$ the FDK is described by formula (\ref{res6}) in a
simplified form, obtained in the case of only two TT spin states
with singlet character (reactive states), $|X_0\rangle = |00\rangle$
and $|X_+ \rangle = \frac{1}{\sqrt{2}}(|\!+\!-\rangle +
|\!-\!+\rangle)$ \cite{Sw2,Bar1}, corresponding to identical
T-excitons (for some additional discussion see also refs [14] and
[15]). This simplified formula can be found, taking into account
that for two reactive spin states (for $N = 2$) the matrix of
weights (\ref{res8}) is represented as ${\hat P}_{\!s}^{} =
\frac{1}{3}(\| X_0\rangle \!\rangle \langle\!\langle X_0 \| + 2\|
X_+\rangle \!\rangle \langle\!\langle X_+ \|)$, and thus in eq.
(\ref{res6}) $\langle\!\langle e\|{\hat P}_{\!s}^{} \hat G
(\epsilon){\hat P}_{\!s}^{}\|e\rangle\!\rangle =
\frac{1}{18}[g_{0}^{}(\epsilon) + 4g_{+}^{}(\epsilon)]$, where
functions $g_{\alpha}(\epsilon),\: (\alpha = 0, +),\,$ are defined
by eq. (\ref{two4b}) with $\zeta_{\alpha}^{} = 0,\: \kappa_{s_{0}} =
k_s^{}/3\,$ and $\,\kappa_{s_{+}} = 2k_s^{}/3$.

The inverse Laplace transformation of thus obtained
$\widetilde{p}_{s}^{} (\epsilon)$ (\ref{res6}) predicts
$p_{s}^{}(t)$ similar to those found for above-mentioned systems
with the same long time dependence ${p}_{s}^{}(t) \sim t_{}^{-3/2}$.
We are not going to thoroughly analyze it, noting only two points
which concern characteristic properties of the FDK, recently
measured in tetracene singlet crystal for $B = 0$ and $B = 8\;{\rm
kG}$ at times $t \lesssim 10_{}^2\; {\rm ns}$ \cite{Bar3}:

a) In general, the behavior of the FDK is reproduced by the proposed
model qualitatively correctly. At intermediate times $t < 20\; {\rm
ns}$ the agreement between experimental and theoretical FDK is
fairly good, as has already been shown above for $B=0$ (Figs 2a,
2b).

b) At longer times, however, some disagreement is observed: the
experimental FDK\cite{Bar3} decreases certainly slower than the
predicted one: ${p}_{s}^{}(t) \sim t_{}^{-3/2}$. The description of
the slower decrease requires some extension of the model, which is a
subject of further investigations (as pointed out below).

\section{Conclusions}

In this work we have proposed the simple and universal model for
studying the kinetics of singlet fission in organic semiconductors
(1). The model enables one to analyze in detail the effect of
three-dimensional diffusive T-exciton migration (in [T+T]-state of
separated T-excitons) on geminate TT-annihilation and thus on
fission process. This model treats the space/time evolution of
TT-pair as transitions between two states: [TT]-state of coupled
T-excitons and [T+T]-state of separated T-excitons, undergoing
three-dimensional relative diffusion [see eqs
(\ref{form2a})-(\ref{form2c})]. The  model quite accurately
describes exponential-type population/depopulation processes in
[TT]-state at relatively short times and small TT-distances, and the
long-time diffusion-like spatial evolution of T-excitons in
[T+T]-state at large TT-distances.

Kinetics of singlet fission is traditionally studied by analyzing
the decay of ${\rm S}_1^{}$-fluorescence intensity ${\bar I}_{_{{\rm
S}_1}}\!(t)$ \cite{Sw1,Sw2}. The proposed model is shown to be able
to describe fairly accurately the normalized FDK ${\bar I}_{_{{\rm
S}_1}}\!(t) = I_{_{{\rm S}_1}}\!(t)/I_{_{{\rm S}_1}}\!(0)$, observed
in the range of times $10^{-1} \: {\rm ns} \lesssim t \lesssim 10^2
\: {\rm ns}$ for a number of systems at magnetic fields $B = 0$ and
$B = 8\:{\rm kG}\,$ \cite{Bar1,Bar4}. Of special interest is the
observed inverse-time dependence ($\sim \xi_{e}^{}t_{}^{-3/2}$) of
the FDK at long times $10\:{\rm ns} \lesssim t \lesssim 10^2 \: {\rm
ns}$, resulted from T-exciton migration. The amplitude of this
dependence is essentially controlled by the parameter $\xi_e^{}$
(\ref{res5}), whose value is determined by characteristic properties
of TT-dissociation and annihilation processes \cite{Shu2,Shu3}.

Concluding our discussion it is worth noting that fairly accurate
description of experimental results demonstrates great
potentialities of the proposed model, which can further be
generalized by taking into consideration characteristic features of
T-exciton spin relaxation and migration in [T+T] state: anisotropy
of migration \cite{Shu2,Shu3}, hopping nature of migration, etc. The
proposed model is expected to be sensitive to details of the
mechanism of T-exciton migration and TT-interaction, and therefore
can be suitable for studying specific structural properties of
organic solids and singlet fission processes in them by analyzing
the FDK in a wide region of times $t \lesssim 10^2\; {\rm ns}$
\cite{Was1,Was2,Bar3}.

\smallskip {\bf Acknowledgements}

The work was supported by the Russian Foundation for Basic Research
(Project 16-03-00052).

\end{document}